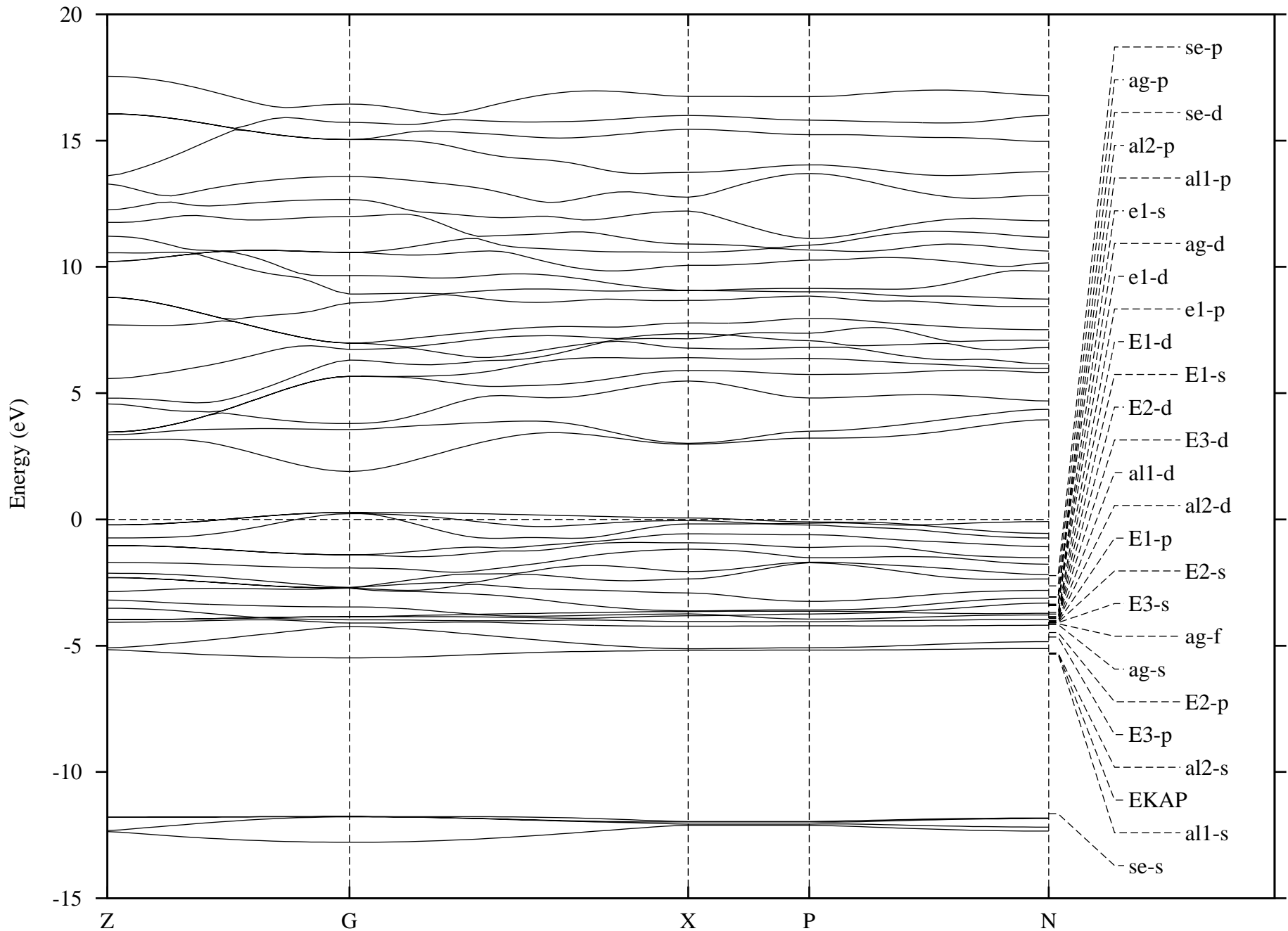

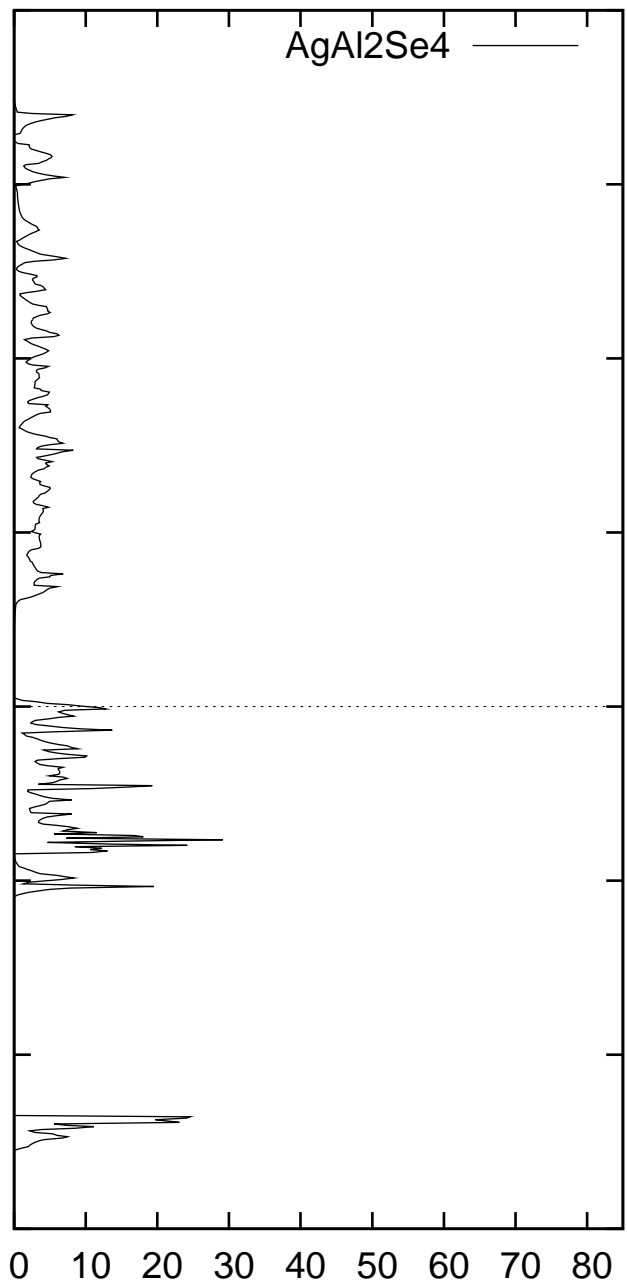

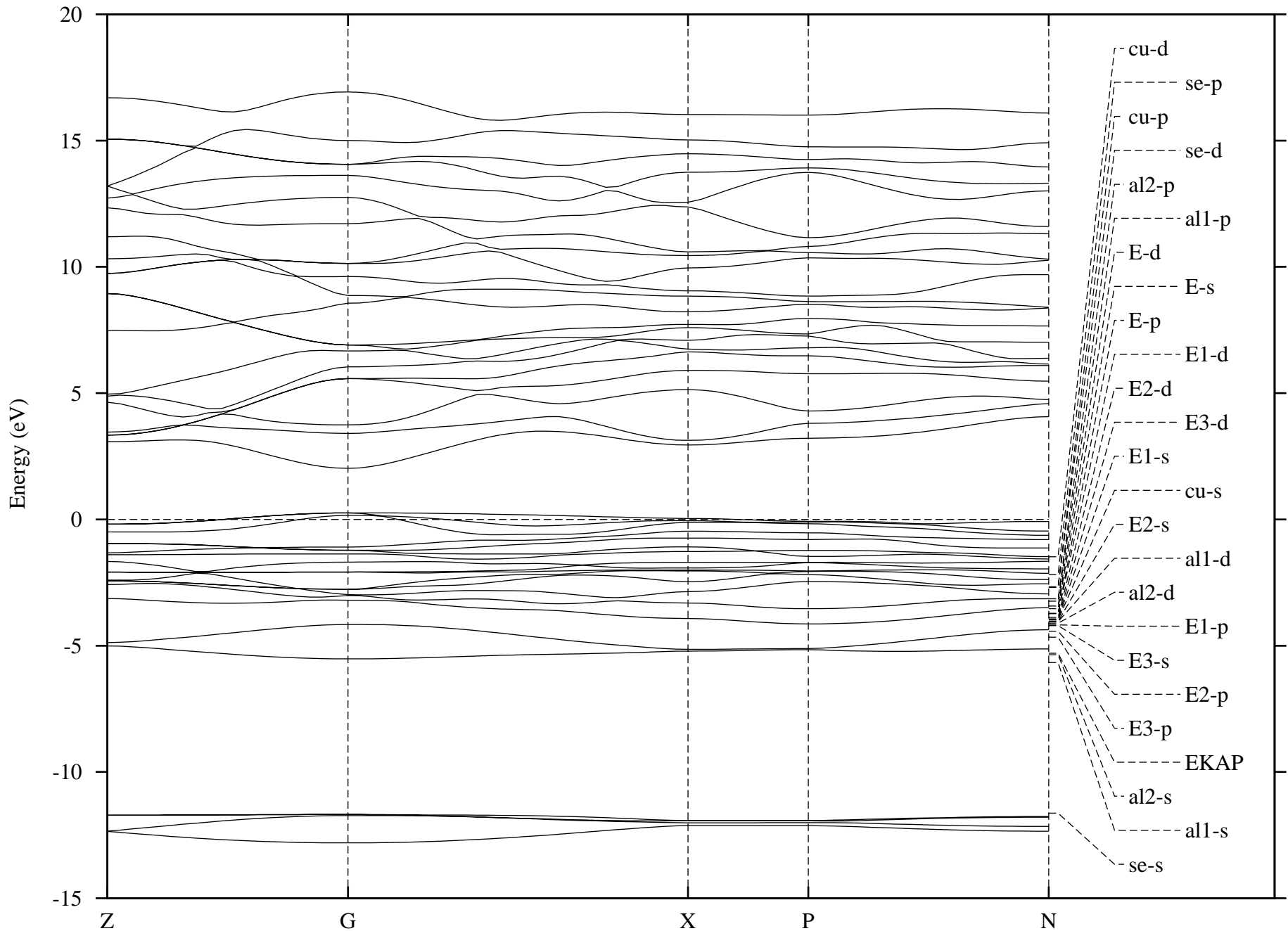

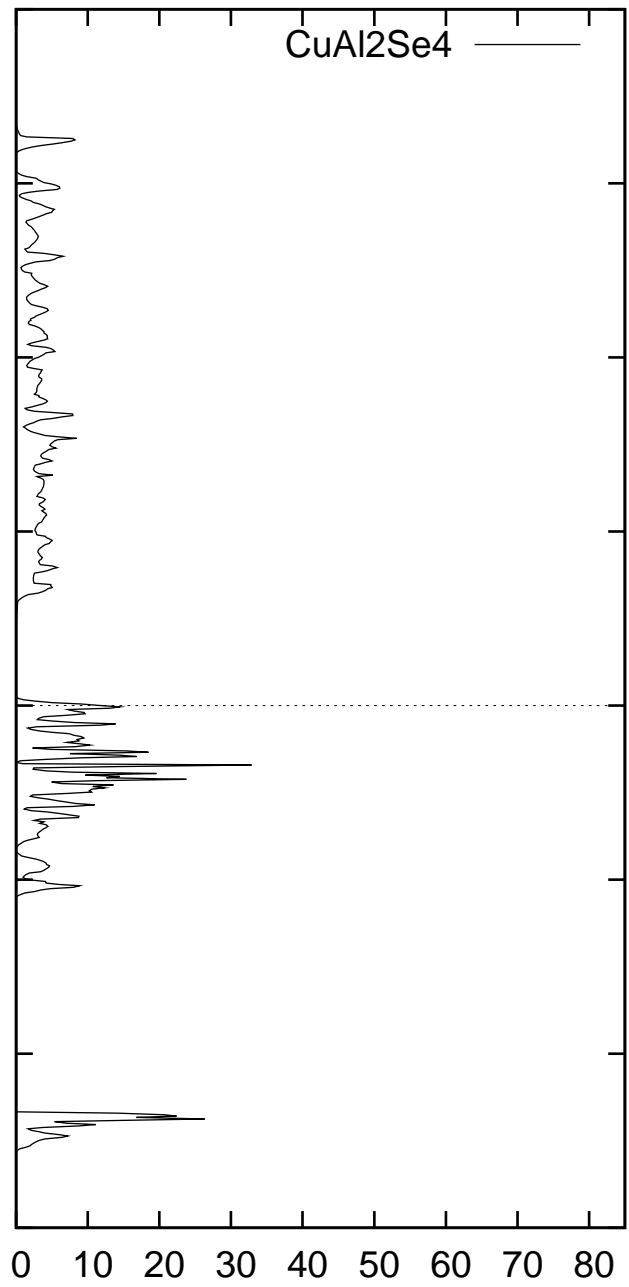

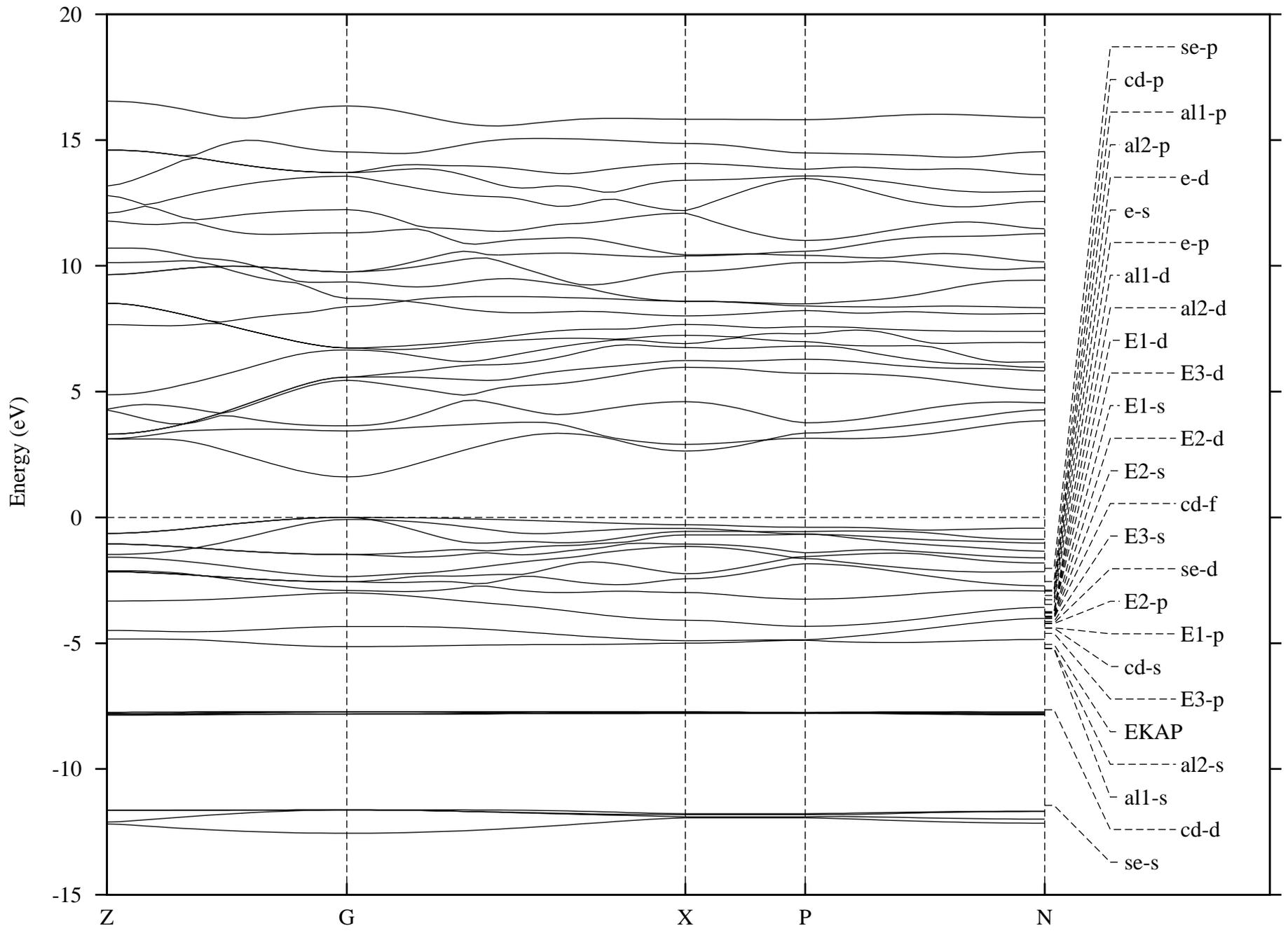

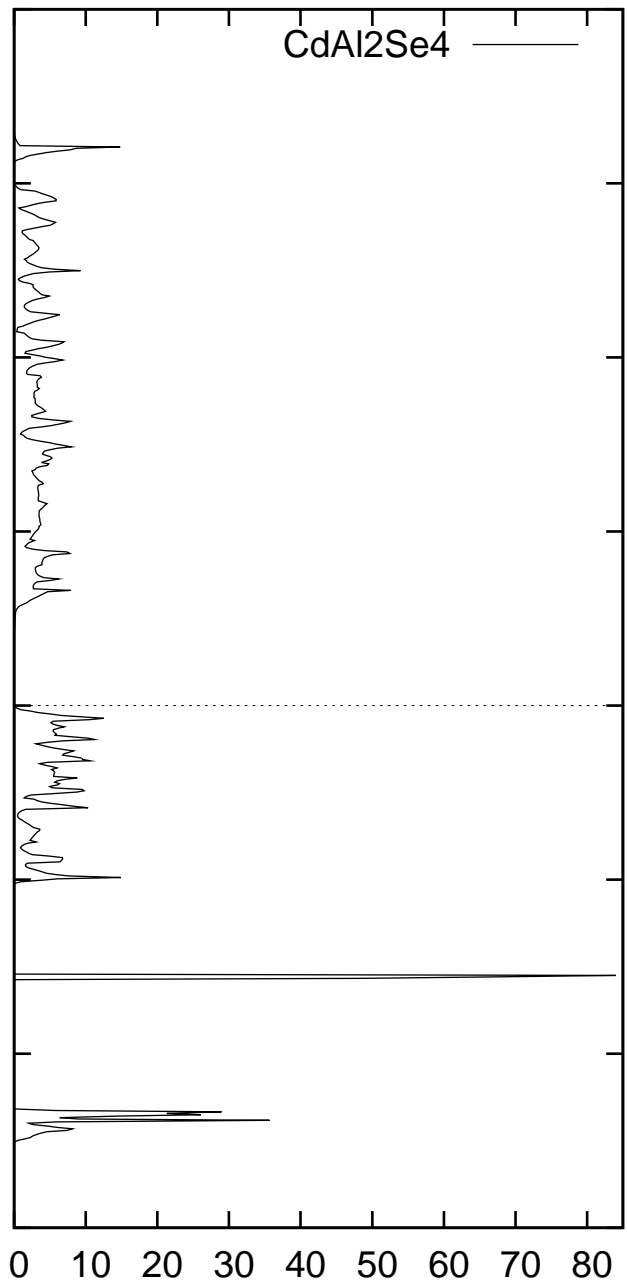

# Electronic and Structural Properties of $AAl_2Se_4 (A = Ag, Cu, Cd, Zn)$ Chalcopyrite Semiconductors

S.Mishra, B.Ganguli[*]

*National Institute of Technology, Rourkela-769008, Odisha, India*

**Abstract**

We have studied the structural and electronic properties of defect chalcopyrite semiconductors $AAl_2Se_4$ (A = Ag, Cu, Cd, Zn) using Density Functional Theory (DFT) based first principle technique within Tight binding Linear Muffin Tin orbital (TB-LMTO) method. Our calculated structural parameters such as lattice constants, anion displacement parameter (u), tetragonal distortion ($\eta$ = c/2a), bond lengths and bulk modulus are in good agreement with other work. Our band structure calculation suggests that these compounds are direct band gap semiconductors having band gaps 2.40, 2.50, 2.46 and 2.82 eV for $AAl_2Se_4(A = Ag, Cu, Cd, Zn)$ respectively. Calculated band gaps are in good agreement with other experimental and theoretical works within LDA limitation. We have made a quantitative estimation of the effect of p-d hybridization and structural distortion on the electronic properties. The reduction in band gap due to p-d hybridization are 19.47%, 21.29%, 0% and 0.7% for $AAl_2Se_4$ (A = Ag, Cu, Cd, Zn) respectively. Increment of the band gap due to structural distortion is 11.62%, 2.45%, 2.92% and 9.30% in case of $AgAl_2Se_4$, $CuAl_2Se_4$, $CdAl_2Se_4$ and $ZnAl_2Se_4$ respectively . We have also discussed the bond nature of all four compounds.

*Keywords:* A. Chalcopyrite; A. Semiconductors; E. Density Functional Theory; E. TB-LMTO

[*]corresponding author. Tel.: +91661 2462725; fax: +91661 2462999
 *Email address:* `biplabg@nitrkl.ac.in` (B.Ganguli)

## 1. Introduction

The ternary semiconducting compounds of type $II - III_2 - VI_4$ belong to one class of functional materials which have attracted great attention due to their potential applications in electro optic, optoelectronic and non-linear optical devices. Few compounds like $CdGa_2Se_4$ and $CdAl_2Se_4$ have already found practical applications such as tunable filters and ultraviolet photodetectors [1, 2]. $HgGa_2S_4$ is considered to be very promising candidate for operating in the mid IR spectral range [3]. Whereas $ZnAl_2Se_4$ is used as a promising material for optoelectronic device application [4]. These defect chalcopyrites have vacancies at the cation sites in such a manner that they do not break translational symmetry. Due to the defect structure the compounds are porous. Large class of defect chalcopyrites and stannite semiconductors have been synthesized [5]. Because of porosity these systems have attracted special attention of the physics community. Various type of impurities including magnetic impurities can be doped into the vacancies to design new class of materials like Dilute magnetic semiconductors (DMS) for spinstronics application [7]. The presence of vacancy and more than two type of atoms provides desired wide band gap, electronic and optical properties to attain a maximum criteria for new emerging functional materials.

In this communication we have investigated the structural and electronic properties of $A - III_2 - VI_4$ compounds where Al is the group III element, Se is group VI element and A represents Ag, Cu, Cd & Zn. There are very few experimental studies carried out and no theoretical study reported for $AgAl_2Se_4$, $CuAl_2Se_4$ and $ZnAl_2Se_4$ [5]. Park et al.[4] studied the optical properties of $ZnAl_2Se_4$ single crystal and found it to be a defect chalcopyrite structure. Extensive experimetal [5, 6, 8, 9] and theoretical studies [10] have been carried out for $CdAl_2Se_4$ compounds. We have chosen this already much studied system to validate our methodology and calculation and extend the study to other systems which have not been much studied. Our main motivation is to study the effect of structure and p-d hybridization on the electronic properties. In our earlier work on pure chalcopyrite semiconductors [11] we have shown that due to the presence of group I element (Cu, Ag), d-orbital contribution is very prominent. This is not valid in case of $II - III_2 - VI_4$ type compounds where group-II elements are Cd and Zn.

For structural properties, we have calculated the lattice parameters, tetragonal distortion, anion displacement parameters and bond lengths by energy minimization proceedure. We have also calculated the bulk modulus using extended Cohen's formula [12]. There have been no calculation so far of bulk modulus of these compounds. We have also found a quantitative relationship between calculated bulk modulus and the lattice parametes. On the basis of bond lengths we have also discussed the nature of bonds in all four defect chalcopyrite compounds. For our study we have used highly successful Density Functional Theory (DFT) based first principle technique, Tight Binding Linearised Muffin-Tin Orbital (TB-LMTO) method. In TB-LMTO method, the basis functions are localized. Therefore, very few basis functions are required to represent the highly localized d-orbital of Ag, Cu, Cd, Zn in the systems under study. Hence the calculation is not only cost effective, it gives also the accurate result.

## 2. Methodology

The ab-initio method is based on Density Functional Theory of Kohn-Sham [13]. The one electron energy is given by Khon-Sham equation.

$$\left[-\nabla^2 + V_{eff}(\mathbf{r})\right]\psi_i(\mathbf{r}) = \varepsilon_\mathbf{i}\psi_\mathbf{i}(\mathbf{r}) \tag{1}$$

where the effective potential,

$$V_{eff}(\mathbf{r}) = 2\int dr' \frac{\rho(\mathbf{r'})}{|\mathbf{r}-\mathbf{r'}|} + 2\sum_R \frac{Z_R}{r-R} + \frac{\delta E_{XC}[\rho]}{\delta\rho(\mathbf{r})} \tag{2}$$

The total electronic energy is a function of electron density which is calculated using variational principle. This requires selfconsistent calculations. In practice the Kohn-Sham orbitals $\psi_i(r)$ are usually expanded in terms of some chosen basis function. We have used the well established TB-LMTO method, discussed in detail elsewhere [14, 15] for the choice of the basis function. Electron correlations are taken within LDA of DFT [13, 16]. We have used the von Barth-Hedin exchange [17] with 512 $\mathbf{k}$-points in the irreducible part of the Brillouin zone. The basis of the TB-LMTO starts from the minimal set of muffin-tin orbitals of a KKR formalism and then linearizes it by expanding around a 'nodal' energy point $E^\alpha_{\nu\ell}$. The wave-function is then expanded in this basis :

$$\Phi_{j\mathbf{k}}(\mathbf{r}) = \sum_\mathbf{L}\sum_\alpha \mathbf{c}^{\mathbf{jk}}_{\mathbf{L}\alpha}\left[\phi^\alpha_{\nu\mathbf{L}}(\mathbf{r}) + \sum_{\mathbf{L'}}\sum_{\alpha'}\mathbf{h}^{\alpha\alpha'}_{\mathbf{LL'}}(\mathbf{k})\dot{\phi}^{\alpha'}_{\nu\mathbf{L'}}(\mathbf{r})\right] \tag{3}$$

where,

$$\begin{aligned}
\phi_{\nu L}^{\alpha}(\mathbf{r}) &= i^{\ell}\, Y_L(\hat{r})\, \phi_{\ell}^{\alpha}(r, E_{\nu\ell}^{\alpha}) \\
\dot{\phi}_{\nu L}^{\alpha}(\mathbf{r}) &= i^{\ell}\, Y_L(\hat{r})\, \frac{\partial \phi_{\ell}^{\alpha}(r, E_{\nu\ell}^{\alpha})}{\partial E} \\
h_{LL'}^{\alpha\alpha'}(\mathbf{k}) &= (C_L^{\alpha} - E_{\nu\ell}^{\alpha})\, \delta_{LL'}\delta_{\alpha\alpha'} + \sqrt{\Delta_L^{\alpha}}\, S_{LL'}^{\alpha\alpha'}(\mathbf{k})\, \sqrt{\Delta_{L'}^{\alpha'}}
\end{aligned}$$

$C_L^{\alpha}$ and $\Delta_L^{\alpha}$ are TB-LMTO potential parameters and $S_{LL'}^{\alpha\alpha'}(\mathbf{k})$ is the structure matrix.

## 3. Result and discussion

*3.1. Structural properties :*

The tetragonal unit cell of a typical chalcopyrite semiconductor consists of two Zinc blende unit cells. There are two types of cations in the unit cell. In a defect chalcopyrite there are 50% vacancy in one type of cation compared to pure chalcopyrite. The vacancies are so created that they dont break the translational symmetry. As shown in figure 1 there are two A atoms, four Al atoms and eight Se atoms in a unit cell. The positions of the various atoms in the tetragonal unit cell are : A : 0.0 0.0 0.0; Vacancy : 0.0 0.5 0.75; B(Al1) : 0.0 0.0 0.5; B(Al2) : 0.0 0.5 0.25; Se: $u_x\ u_y\ u_z$ where '$u_x$', '$u_y$' and '$u_z$' are anion displacement parameters along three axes. For ideal case, $\eta(c/2a) = 1$ and $u_x$, $u_y$ and $u_z$ are 0.25, 0.25 and 0.125 respectively. In all four systems each Se atom has one A type cation, two Al-cations and one vacancy as nearest neighbors as shown in figure 1. Due to different atoms and one vacancy as neighbors the Se atom acquires an equilibrium position closer to the vacancy than to the other three cations. This new position of the anion is called anion-displacement. Se atoms shifts along all the three directions unlike only along x-direction as found in case of non-defect chalcopyrites [11, 18]. This is due to the reduction in symmetry in case of defect system. Therefore all the three cations-Se bond lengths are inequivalent.

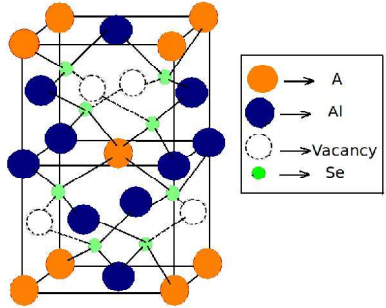

Figure 1: One unit cell of defect chalcopyrite semiconductor $AAl_2Se_4$ (A = Ag, Cu, Cd, Zn)

Table 1: Calculated structural parameters

| Compounds | a (Å) | c/a | $a_{exp}$ (Å) | $c/a_{exp}$ | $u_x$ | $u_y$ | $u_z$ | B (GPa) |
|---|---|---|---|---|---|---|---|---|
| $AgAl_2Se_4$ | 5.72 | 1.979 | | | 0.262 | 0.248 | 0.130 | 46.87 |
| $CuAl_2Se_4$ | 5.65 | 1.998 | | | 0.271 | 0.251 | 0.130 | 47.98 |
| $CdAl_2Se_4$ | 5.76 | 1.996 | $5.73^a$ | $1.85^a$ | 0.264 | 0.249 | 0.132 | 44.89 |
| $ZnAl_2Se_4$ | 5.53 | 1.981 | $5.49^a$ | $1.98^a$ | 0.266 | 0.247 | 0.131 | 52.06 |

$^a$ Ref.[5]

For self consistent calculation, we introduce empty spheres because the packing fraction is low due to tetrahedral co-ordination of ions. We ensure proper overlap of muffin tin spheres for self consistency and the percentage of overlap is found. Table 1 shows the calculated structural parameters 'a', 'c', tetragonal distortion, anion displacement and bulk modulus (B). These parameters are found by energy minimization procedure. We have calculated the bulk modulus 'B' using extended Cohen formula [12] for $II-III_2-VI_4$ compounds.

$$B = \frac{1971 - 220\lambda}{4} \sum_{i=1,2,3} \frac{1}{d_i^{3.5}} \qquad (4)$$

where B is in GPa and the nearest-neighbor distances $d_i$ in $A^0$. The ionicity coefficient $\lambda$ is taken equal to 2, analogous to II-VI semiconductors. The calculated result shows an inverse proportionality relation between bulk modulus and lattice constant 'a'. Our result agrees with the similar study done for semiconductors like for II-VI type semiconductors [19].

*3.2. Electronic properties*

(i) $AgAl_2Se_4$ : The band structure and total density of states (TDOS) (figure 2) show that this compound is a slightly p-type semiconductor. There are three major sub valence bands of different band widths. The first two subbands of band widths 4.5 eV and 1.1 eV respectively below the valence band maxima are separated by very narrow band gap of 0.1 eV. The lowest subband having band width 1.0 eV is formed mainly due to the contribution of Se-4s states. There is a large band gap of $\simeq 6.2 eV$ between the lowest and second subbands. The second subband is formed due to the admixture of Al-s and Se-p orbitals. Figure 3 shows Partial density of states (PDOS) for Ag-d, Se-p, Al-s and Al-p. It is clear from the figure that Ag-d and Se-p hybridization contribute to upper valence band near fermi level and there is a very weak contribution of Ag-d states to conduction band. The main contribution to conduction band is due to Al-p and Se-p states and very weak contribution from Al-s orbitals. The conduction band width is $\simeq 14.4$ eV.

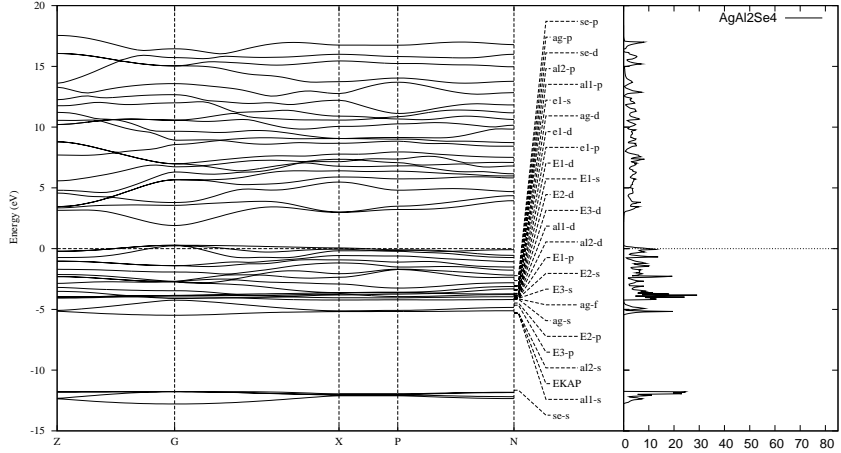

Figure 2: Band structure and TDOS for non-ideal $AgAl_2Se_4$

(ii) $CuAl_2Se_4$ : Like $AgAl_2Se_4$, this compound is also slightly p-type semiconductor. The band structure and TDOS (figure 4), show three sub valence bands of different band widths. The first two subbands having band widths 4.4 eV and 1.3 eV respectively are separated by very narrow band gap of 0.1 eV. The lowest subband having band width 1.2 eV is formed due to Se-4s states. There is a large band gap of nearly 6.1 eV between the lowest and just above subbands. The second subband is mainly formed due to the admixture of Al-s and Se-p states. The upper valence band is dominated by Cu-d and Se-p hybride orbitals. The conduction band width is approxi-mately 14.0 eV. Calculated PDOS for s, p, d orbitals of Cu, Al and se (figure 5) show that the conduction band has main contribution from Al-p and Se-p and a very weak contribution from Al-s states.

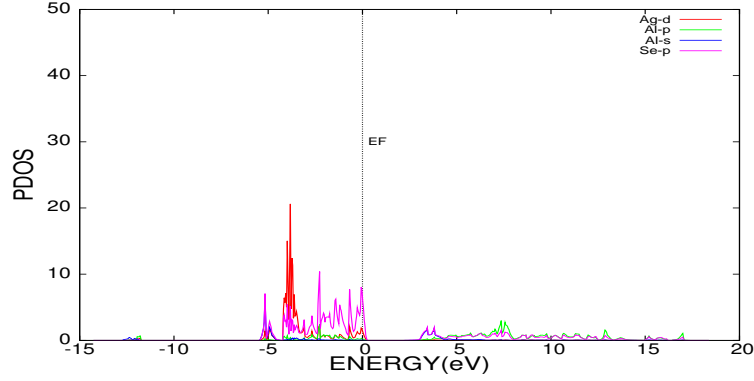

Figure 3: PDOS of Ag-d, Al-p, Al-s and Se-p for non-ideal $AgAl_2Se_4$

(iii) $CdAl_2Se_4$ : Unlike in $AgAl_2Se_4$ and $CuAl_2Se_4$, this compound does not behave as a p-type semiconductor. The band structure and TDOS (figure 6) shows three major subvalence bands of different band widths. The two upper most subvalence bands have band widths of 5.1 eV and 0.2eV respectively. They are separated by 2.6 eV. The second band is mainly formed due to the contribution of Cd-d and very weak contribution from Se-p states. The lowest band of band width 0.9 eV is formed due to the contribution of Se-4s states. Figure 7 for PDOS shows that the middle valence subband is formed due to the Cd-d orbitals. There is no contribution of Cd-d orbital to conduction band. Unlike in $AgAl_2Se_4$ and $CuAl_2Se_4$, for this compound the main contribution to upper most valence band is due to the contribution of the admixture of Se-p, Al-s and very weak Al-p states. The main contribution to conduction band is from Se-p, Al-p states. The conduction band width is found to be 14.0 eV.

(iV) $ZnAl_2Se_4$ : The band structure and TDOS (figure 8) show three major sub valence bands. The first two subbands having band widths 5.5 eV and 0.2 eV respectively are separated by band gap of 1.0 eV . The lowest subband having band width 1.3 eV is formed mainly due to the contribution of Se-4s states. There is a large band gap of $\simeq 5.1$eV between the lowest and the second subband. The second subband is formed due to the contribution of Zn-d and there is very weak contribution of Se-p orbital. PDOS for Zn-d, Al-p Al-s and Se-p are shown in figure 9. From this figure it is clear that

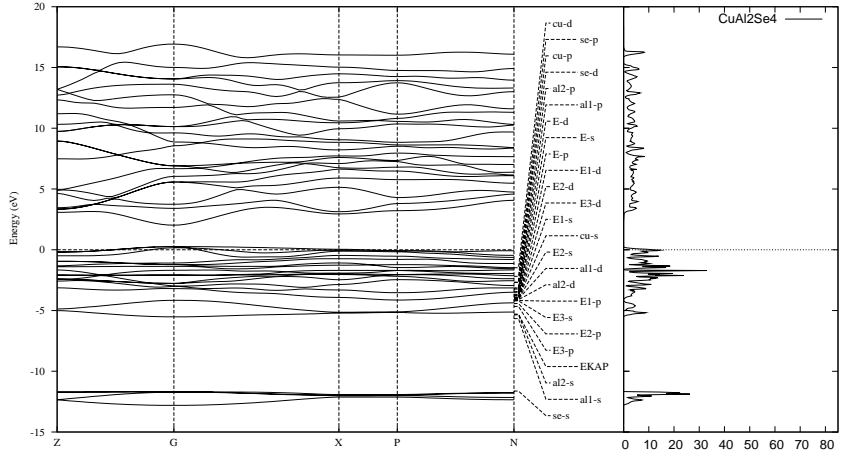

Figure 4: Band structure and TDOS for non-ideal $CuAl_2Se_4$

Table 2: Energy gaps of defect chalcopyrites

| Compounds | Experiment (eV) | Other calculations (eV) | Our work (eV) |
|---|---|---|---|
| $AgAl_2Se_4$ | | $2.16^b$ | 2.40 |
| $CuAl_2Se_4$ | $2.65$-$3.02^a$ | $2.49^b$ | 2,50 |
| $CdAl_2Se_4$ | $3.07^c$ | $3.54^b$ | 2.46 |
| $ZnAl_2Se_4$ | $3.52^d$ | $3.65^b$ | 2.82 |

[a] Ref.[5]  [b] Ref.[10]  [c] Ref.[6]  [d] Ref.[4]

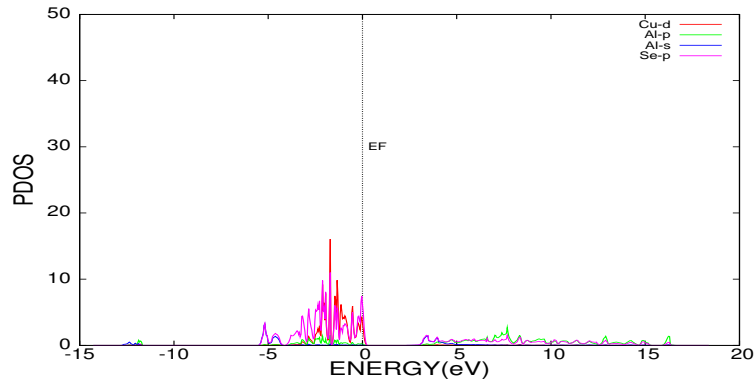

Figure 5: PDOS of Cu-d, Al-p, Al-s and Se-p for non-ideal $CuAl_2Se_4$

Zn-d orbital do not contribute to upper valence band near fermi level. The contribution to conduction band is due to Al-p, Al-s and Se-p states. The conduction band width is $\simeq$ 14.4 eV.

In all the cases valence band maximum (VBM) and conduction band minimum (CBM) are located at center of Brillouin zone denoted as 'G' ($\Gamma$ point). This indicates that they are all direct band gap compounds. Experimental, other theoretical and our calculated band gaps are listed in table 2. It is known that LDA underestimates band gap by 30%. If we correct this error, our results are in good agreement with experimental band gap. No experimental study of band gap for $AgAl_2Se_4$ is reported in literature. Jiang et.al. [10] have made a correction to LDA using scisser effect for a series of defect chalcopyrites by raising the energy at symmetric points in the band structure. But the correction overestimate the experimental band gaps for $ZnAl_2Se_4$ and $CdAl_2Se_4$.

*3.3. Effect of p-d hybridization on electronic properties*

It is known that p-d hybridization has significant effect on the band gap in the case of Ag and Cu based compounds [11, 20]. To see this effect explicitly, we have calculated the band structure and TDOS without the contribution of the d-orbitals for ideal $AAl_2Se_4$ systems. Therefor we have first freezed the d-electrons and have treated these electrons as core electrons. Figures 10 respectively show the TDOS with d-electron of A atoms as frozen for $AAl_2Se_4$. We have summarized the band gaps with and without contribution of d-electrons of Ag, Cu, Cd and Zn in table 3. The calculated result shows that there is a significant reduction of band gaps particularly in case of

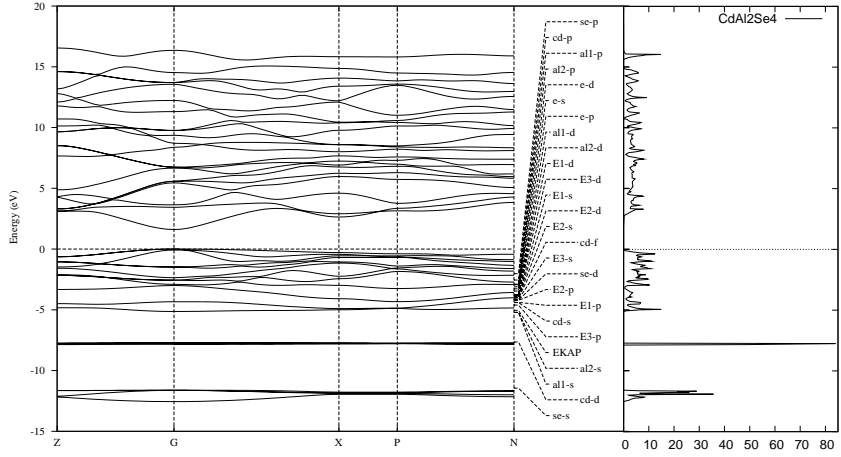

Figure 6: Band structure and TDOS for non-ideal $CdAl_2Se_4$

Table 3: % of Reduction in band gap(eV) due to hybridization for ideal case.

| Systems | With hybridization | Without hybridization | Band gap reduction |
|---|---|---|---|
| $AgAl_2Se_4$ | 2.15 | 2.67 | 19.47% |
| $CuAl_2Se_4$ | 2.44 | 3.10 | 21.29% |
| $CdAl_2Se_4$ | 2.39 | 2.39 | 0% |
| $ZnAl_2Se_4$ | 2.58 | 2.60 | 0.7% |

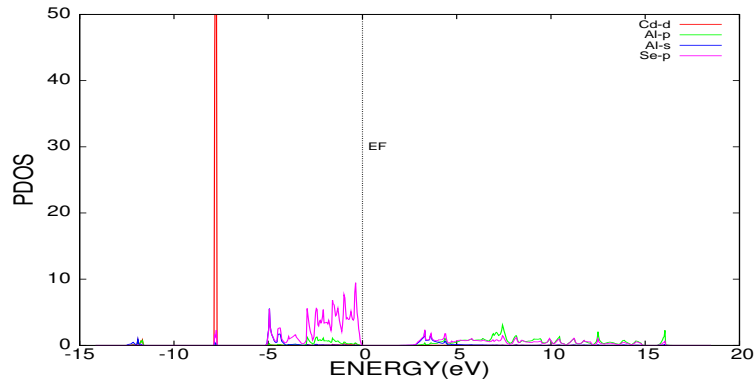

Figure 7: PDOS of Cd-d, Al-p, Al-s and Se-p for non-ideal $CdAl_2Se_4$

$AgAl_2Se_4$ and $CuAl_2Se4$ . The reduction is 19.47% for $AgAl_2Se_4$, 21.29% for $CuAl_2Se_4$, $\sim$ 0% for $CdAl_2Se_4$ and 0.7% for $ZnAl_2Se_4$. The p-d hybridization in general chalcopyrite semiconductors can be interpreted on the basis of simple molecular orbital considerations [20]. The p-orbitals that possess the $\Gamma_{15}$ symmetry hybridize with those of the d-orbitals that present the same symmetry. This hybridization forms a lower bonding state and an upper antibonding state. The antibonding state that constitutes the top of the valence band is predominantly formed by higher energy anion p-states and the bonding state is constituted by the lower energy cation d-states. Perturbation theory [21] suggests that the two states $\Gamma_{15}(p)$ and $\Gamma_{15}(d)$ will repel each other by an amount inversely proportional to the energy difference between p and d states. So this raising of the upper most state causes a gap reduction. But in defect $AgAl_2Se_4$ and $CuAl_2Se_4$ there is a reduction in the atomic percentage of Ag and Cu relative to that $AgAlSe_2$ and $CuAlSe_2$. So the repulsion between $\Gamma_{15}(p)$ and $\Gamma_{15}$ d) decreases and the antibonding state is depressed downwards leading to an increase in band gap in case of $AgAl_2Se_4$ and $CuAl_2Se_4$. Thus all the Ag and Cu-defficient defect chalcopyrites have band gaps greater than that the corresponding pure chalcopyrites [11]. The p-d hybridization in Cu-based chalcopyrites is known to contribute more to band gap reduction than Ag-based. We find that when cation atomic size increases, the band gap always decreases. This is in agreement with other work [22].

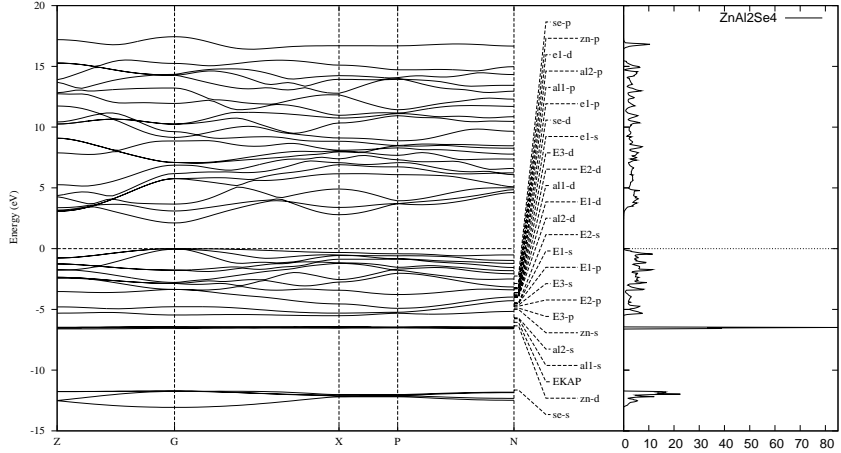

Figure 8: Band structure and TDOS for non-ideal $ZnAl_2Se_4$

Table 4: Effect of structural distortion on band gap(eV).

| Systems | Ideal | Non-ideal | Increment in band gap |
|---|---|---|---|
| $AgAl_2Se_4$ | 2.15 | 2.40 | 11.62% |
| $CuAl_2Se_4$ | 2.44 | 2.50 | 2.45% |
| $CdAl_2Se_4$ | 2.39 | 2.46 | 2.92% |
| $ZnAl_2Se_4$ | 2.58 | 2.82 | 9.30% |

*3.4. Structural effect on electronic properties*

Table 4 shows calculated band gaps for ideal and non-ideal structures. It shows that the structural distortion like bond alternation and tetragonal distortion have significant contribution in the band gap. A close comparision of TDOS for ideal (figure 11) and non-ideal (figure 2) cases of $AgAl_2Se_4$ shows distinct differences in the structure in DOS. For example a sharp peak is found nearly at energy -4.0 eV for ideal $AgAl_2Se_4$ compared to the corresponding non-ideal case. The sharp peak comes due to the contribution of Ag-d orbitals. This shows that structural distortion not only increase the band gap but it has significant effect on overall electronic properties as well. Similar results are also found for $CuAl_2Se_4$, $CdAl_2Se_4$ and $ZnAl_2Se_4$ systems. In these systems also the sharp peaks come due to Cu-d, Cd-d and Zn-d orbitals respectively. There are effects on conduction band also in all four defect chalcopyrites. The effect of distortion on valence and conduc-tion bands show that structural distortion is also responsible for significant change in optical properties of such semiconductors.

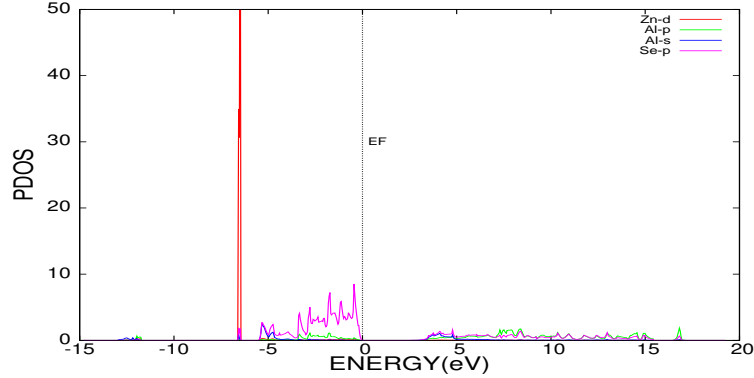

Figure 9: PDOS of Zn-d, Al-p, Al-s and Se-p for non-ideal $ZnAl_2Se_4$

*3.5. Bond nature*

Calculated bond lengths and corresponding covalent and ionic radii are listed in table 5. The calculated values of two Al-Se bond lengths are corresponding to the two inequivalent sites for Al. We observe that our calculated Ag-Se, Al1-Se and Al2-Se bond lengths are closer to the sum of the covalent radii rather than the sum of the ionic radii. Similarlly in the case of $CuAl_2Se_4$ and $CdAl_2Se_4$ all the three calculated bond lengths are closer to the sum of the covalent radii of the atoms than the sum of the ionic radii. But in the case of $ZnAl_2Se_4$ the bond length of Al1-Se is closer to the sum of the ionic radii of Al and Se than the sum of the covalent radii. This shows that all bonds except Al1-Se for $ZnAl_2Se_4$ are covalent in nature. This little ionicity of Al1-Se bond in $ZnAl_2Se_4$ increases the band gap in comparision to other three compounds.

## 4. Conclusion

Calculations and study of $AAl_2Se_4$ (A = Ag, Cu, Cd, Zn) suggest that these compounds are direct band gap semiconductors with band gaps of 2.40V, 2.50 eV, 2.46 and 2.82 eV respectively. Our study further shows that electronic properties of these semiconductors significantly depend on the type of hybridization, structural distortion and the bond nature of the compounds. The calculation is carried out using DFT based TB-LMTO method. We have used LDA for our exchange co-relation functional. Taking into account of the underestimation of band gap by LDA, our result of band gap and structutal properties agree with experimental values. Detail study of TDOS and PDOS shows that p-d hybridization between atom A-d and anion-p reduces the band gap. The reduction is 19.47%, 21.29%, 0% and 0.7% respectively for A = Ag, Cu, Cd and Zn. Increment of the band gap due to structural distortion is 11.62%, 2.45%, 2.92% and 9.30% in case of $AgAl_2Se_4$, $CuAl_2Se_4$, $CdAl_2Se_4$ and $ZnAl_2Se_4$ respectively. The bond nature of these compounds are also discussed.

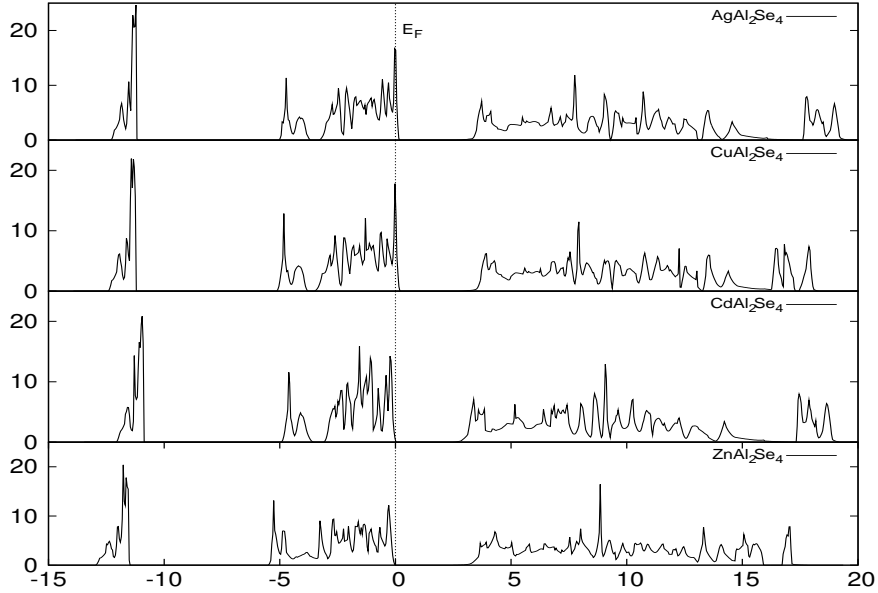

Figure 10: TDOS vs ENERGY(eV) plot for ideal and without hybridization for $AAl_2Se_4$

## Acknowledgement

This work was supported by Department of Science and Technology, India, under the grant no.SR/S2/CMP-26/2007. We would like to thank Prof. O.K. Andersen, Max Planck Institute, Stuttgart, Germany, for kind permission to use the TB-LMTO code developed by his group.

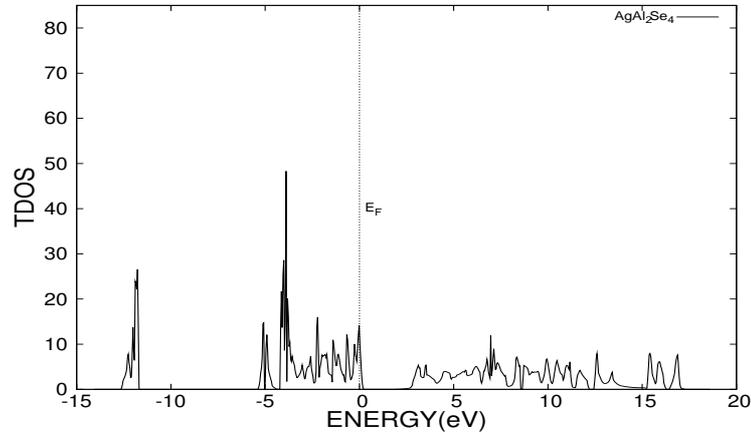

Figure 11: TDOS for ideal $AgAl_2Se_4$

Table 5: Calculated bond lengths in Å.

| Systems | $R_{A-Se}$ (Å) | $R_{Al1-Se}$ (Å) | $R_{Al2-Se}$ (Å) | $R_{Vacancy-Se}$ (Å) |
|---|---|---|---|---|
| $AgAl_2Se_4$ | 2.533 (2.65)* [2.92]** | 2.468 | 2.483 | 2.406 |
| $CuAl_2Se_4$ | 2.551 (2.52)* [2.71]** | 2.409 | 2.481 | 2.349 |
| $CdAl_2Se_4$ | 2.583 (2.64)* [2.93]** | 2.498 | 2.498 | 2.396 |
| $ZnAl_2Se_4$ | 2.467 (2.42)* [2.72]** | 2.385 | 2.412 | 2.288 |

* Sum of the covalent radii. ** Sum of the ionic radii.
Sum of the covalent radii of Al-Se is 2.41(Å).
Sum of the ionic radii of Al-Se is 2.37(Å).

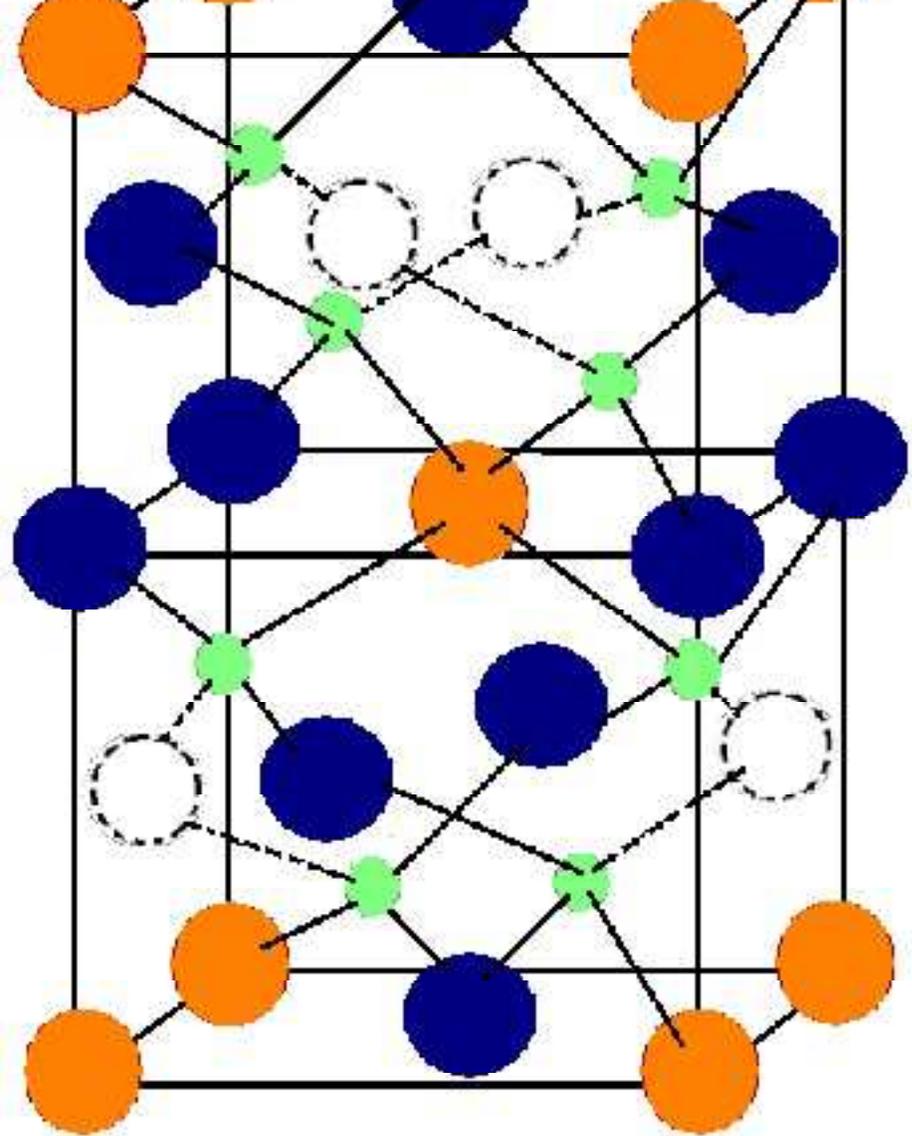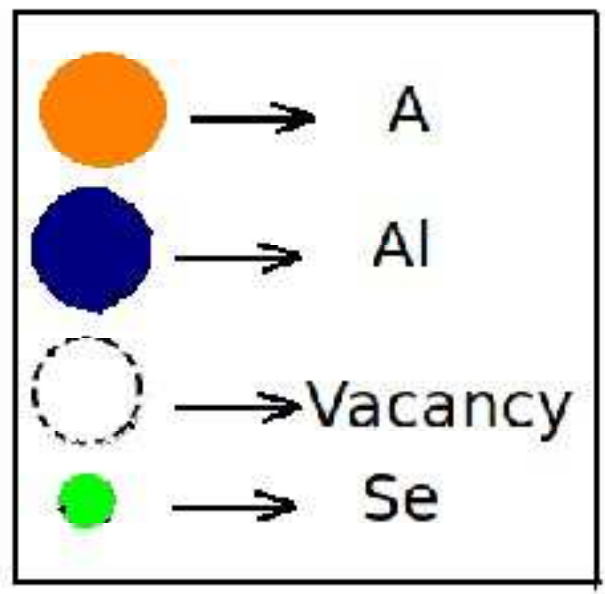

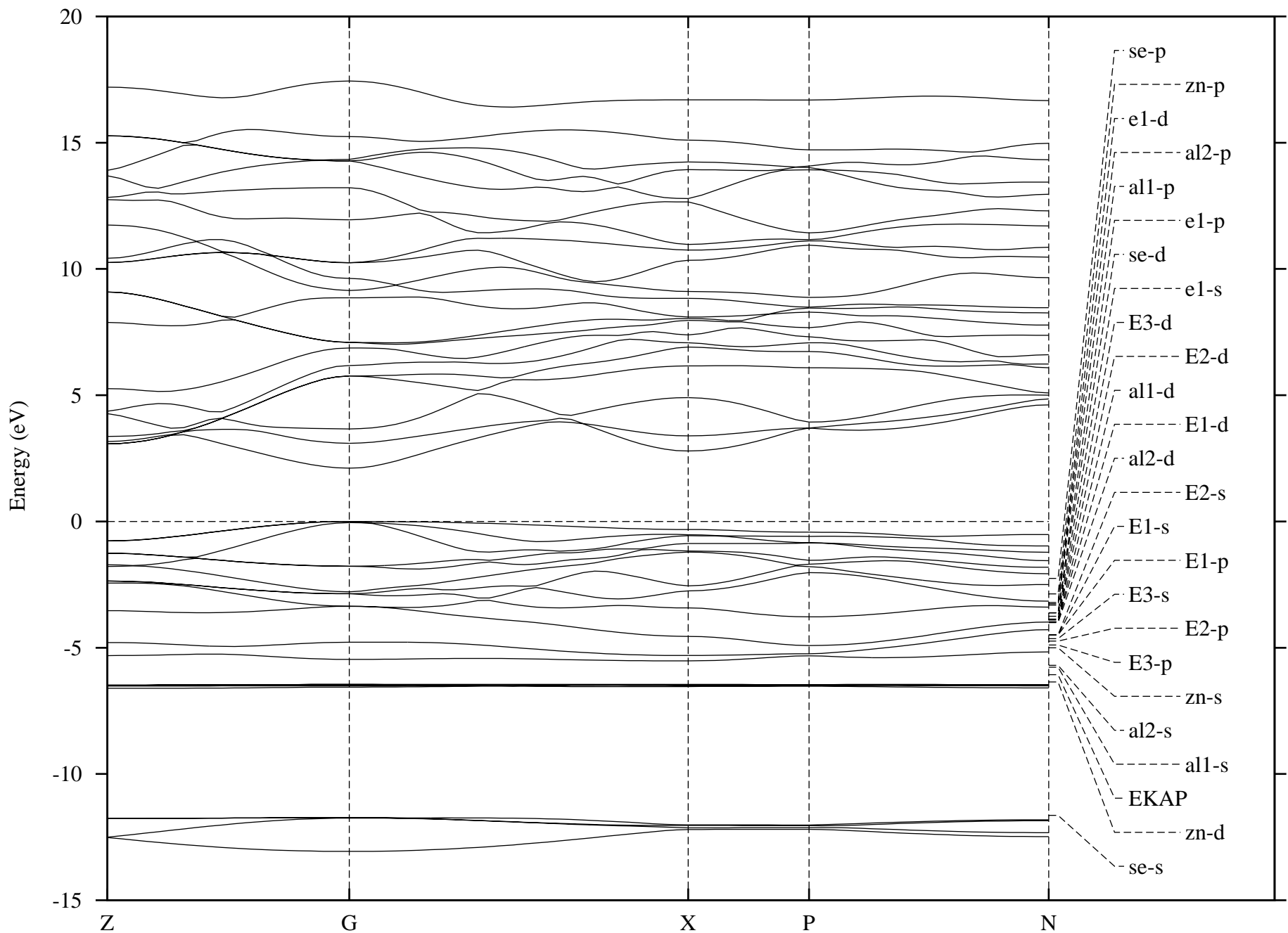

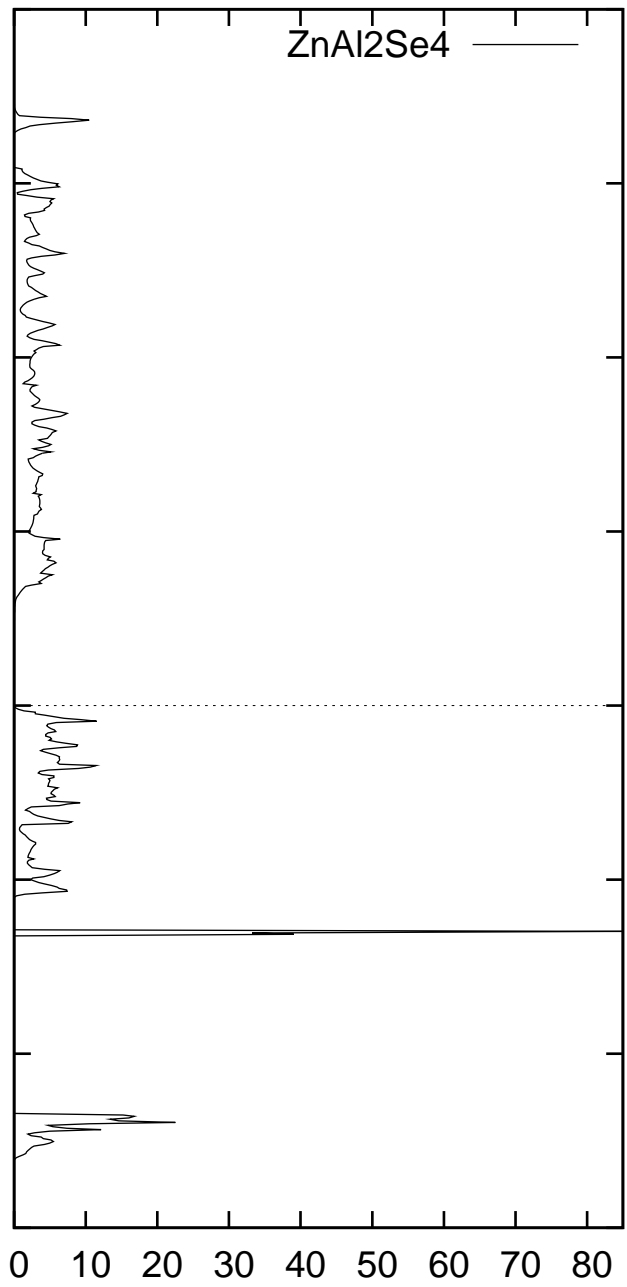

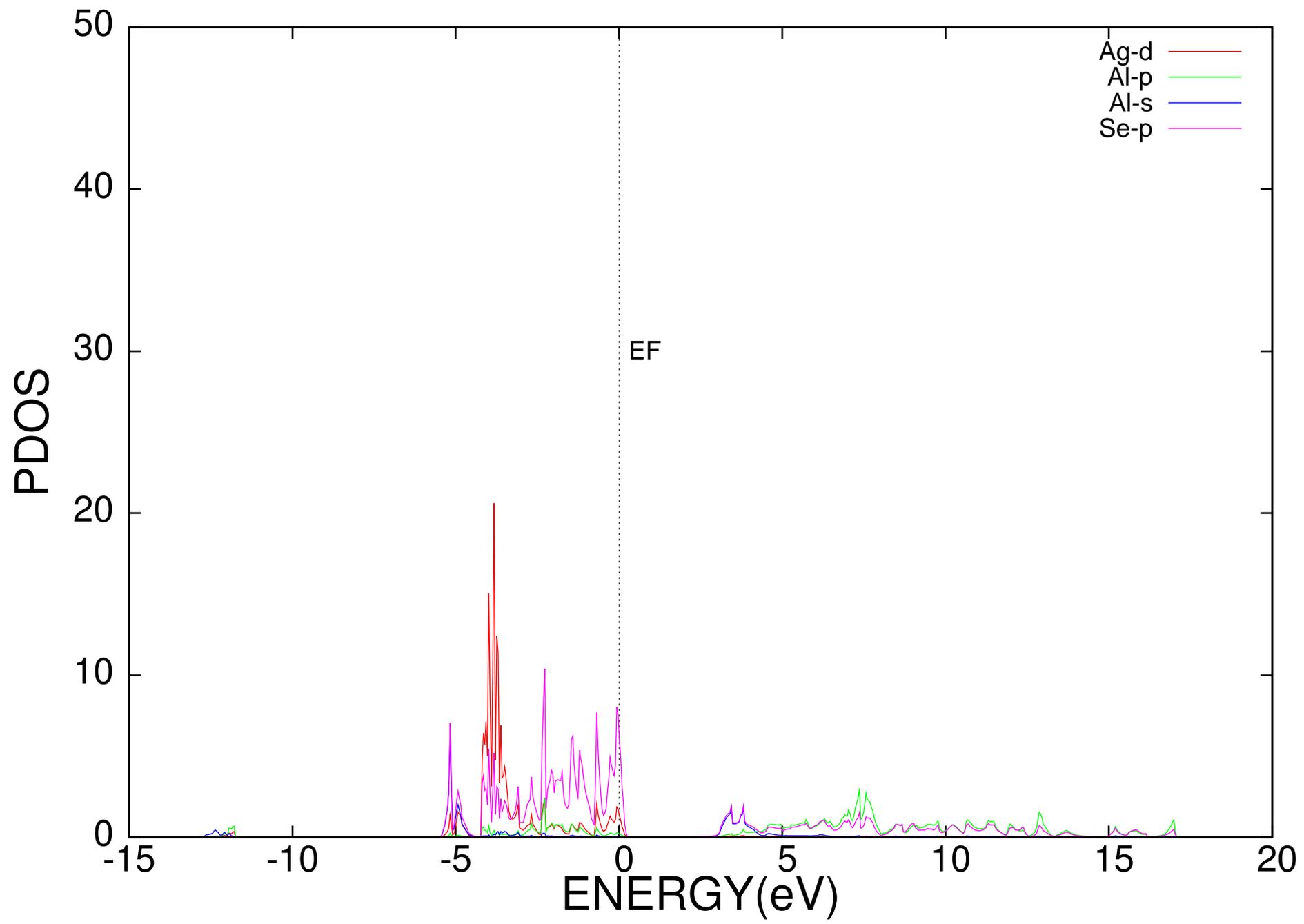

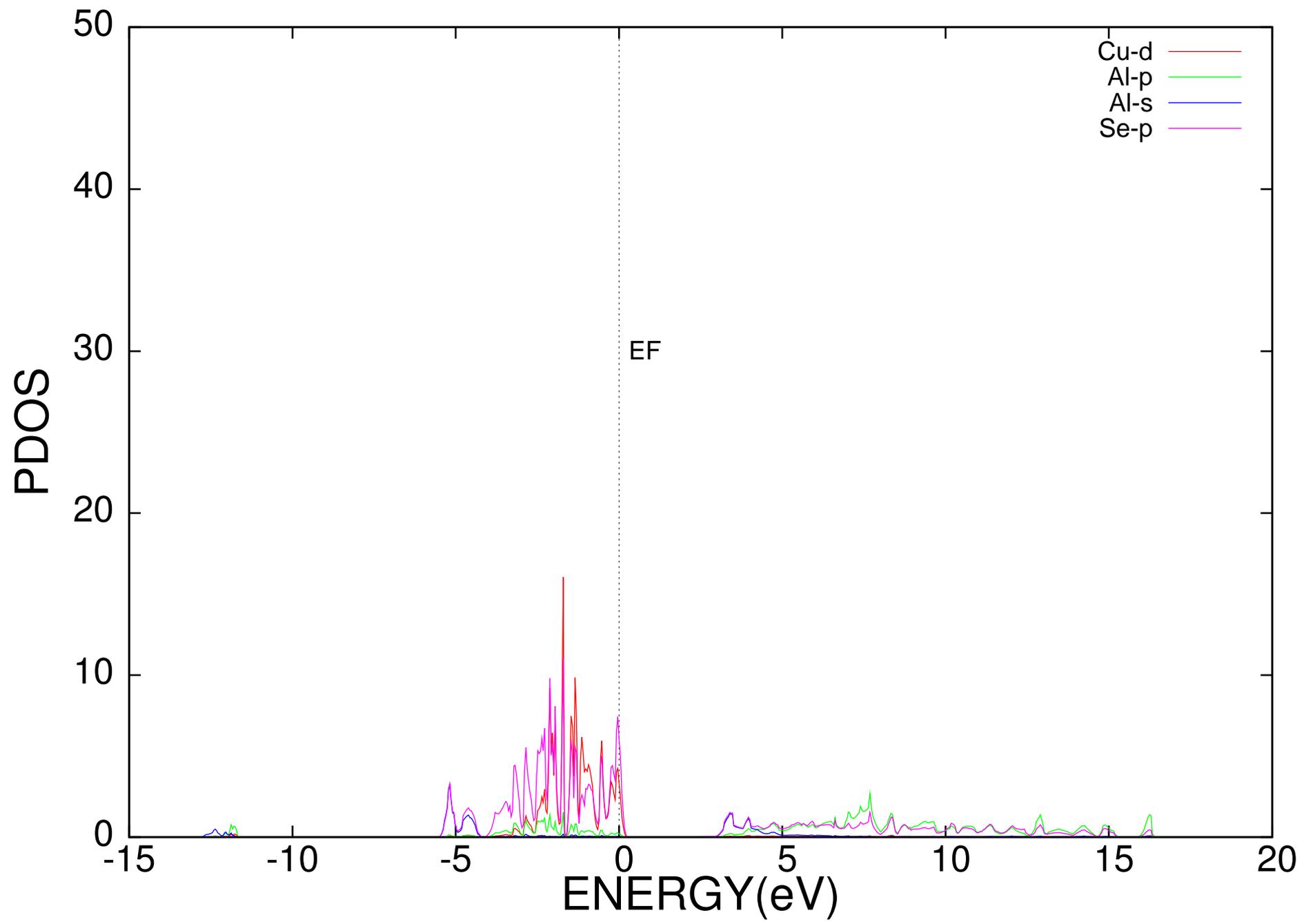

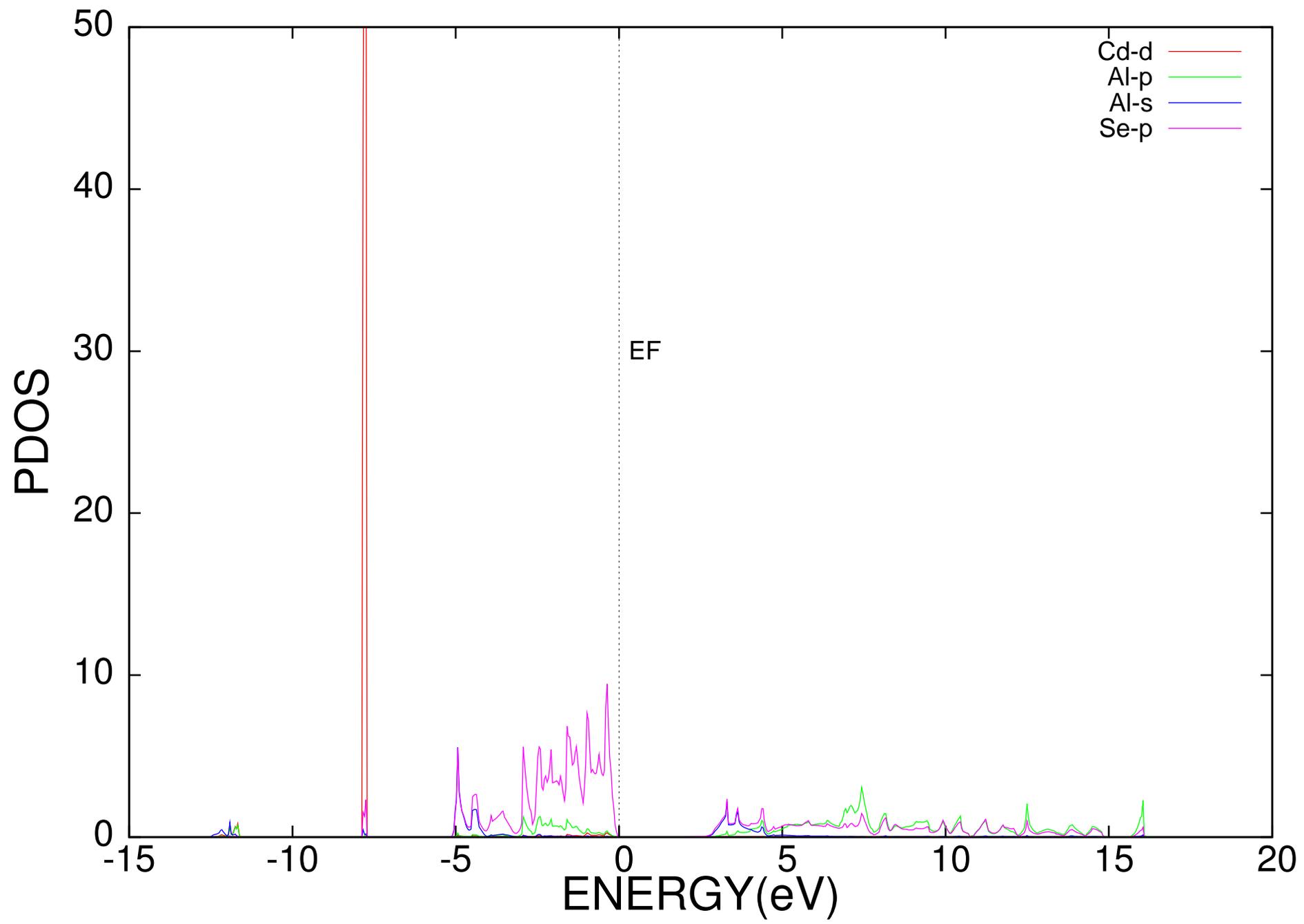

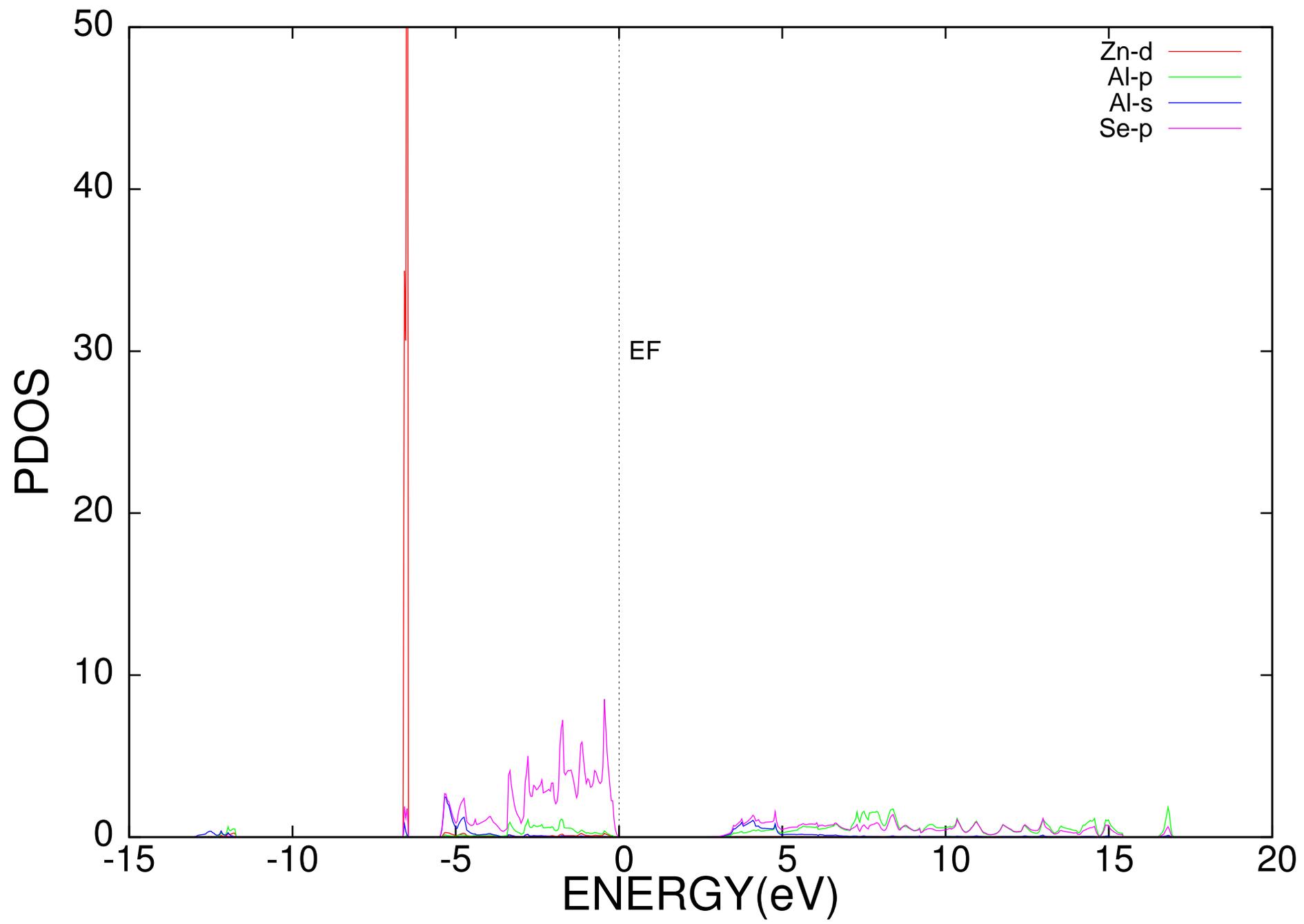

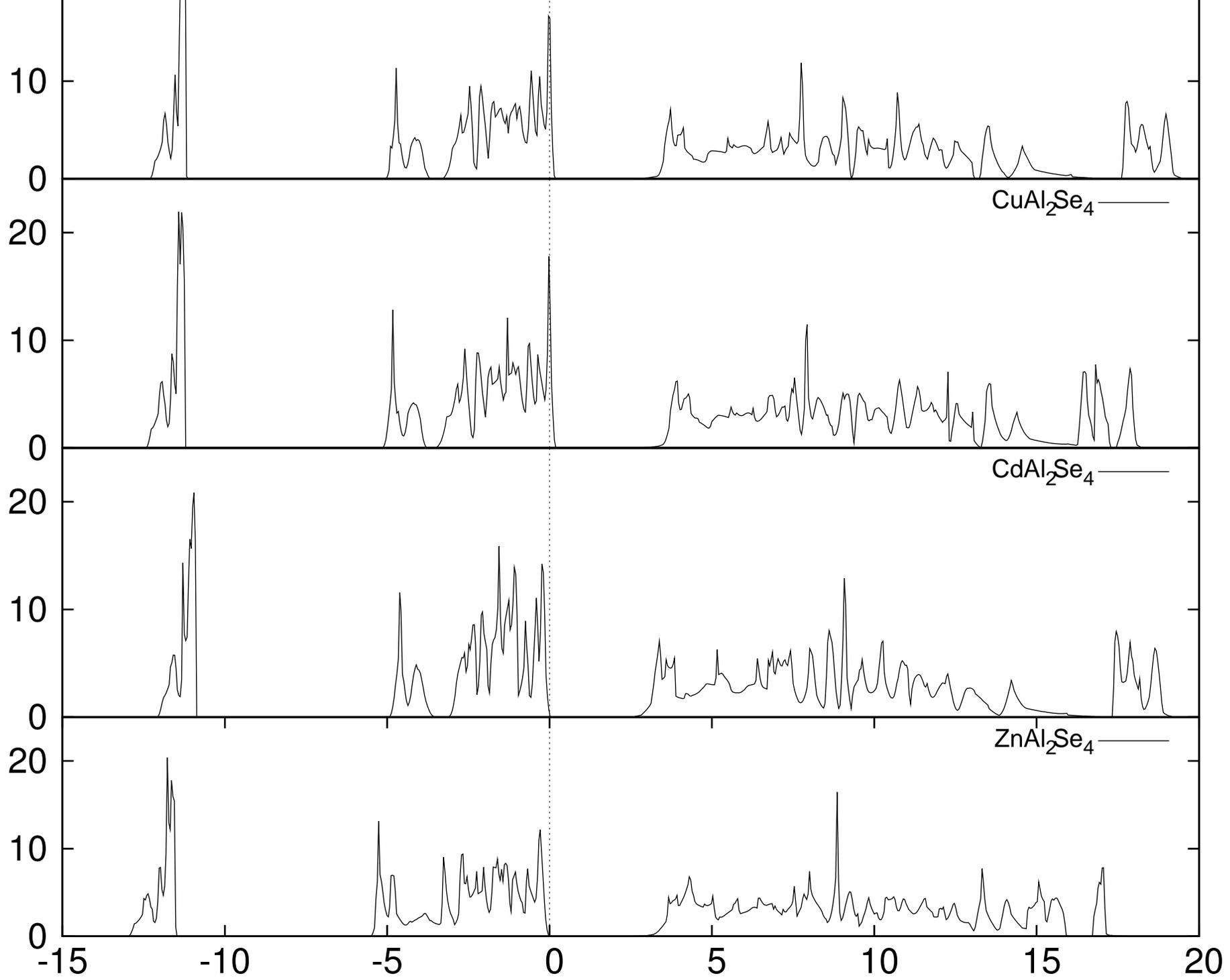

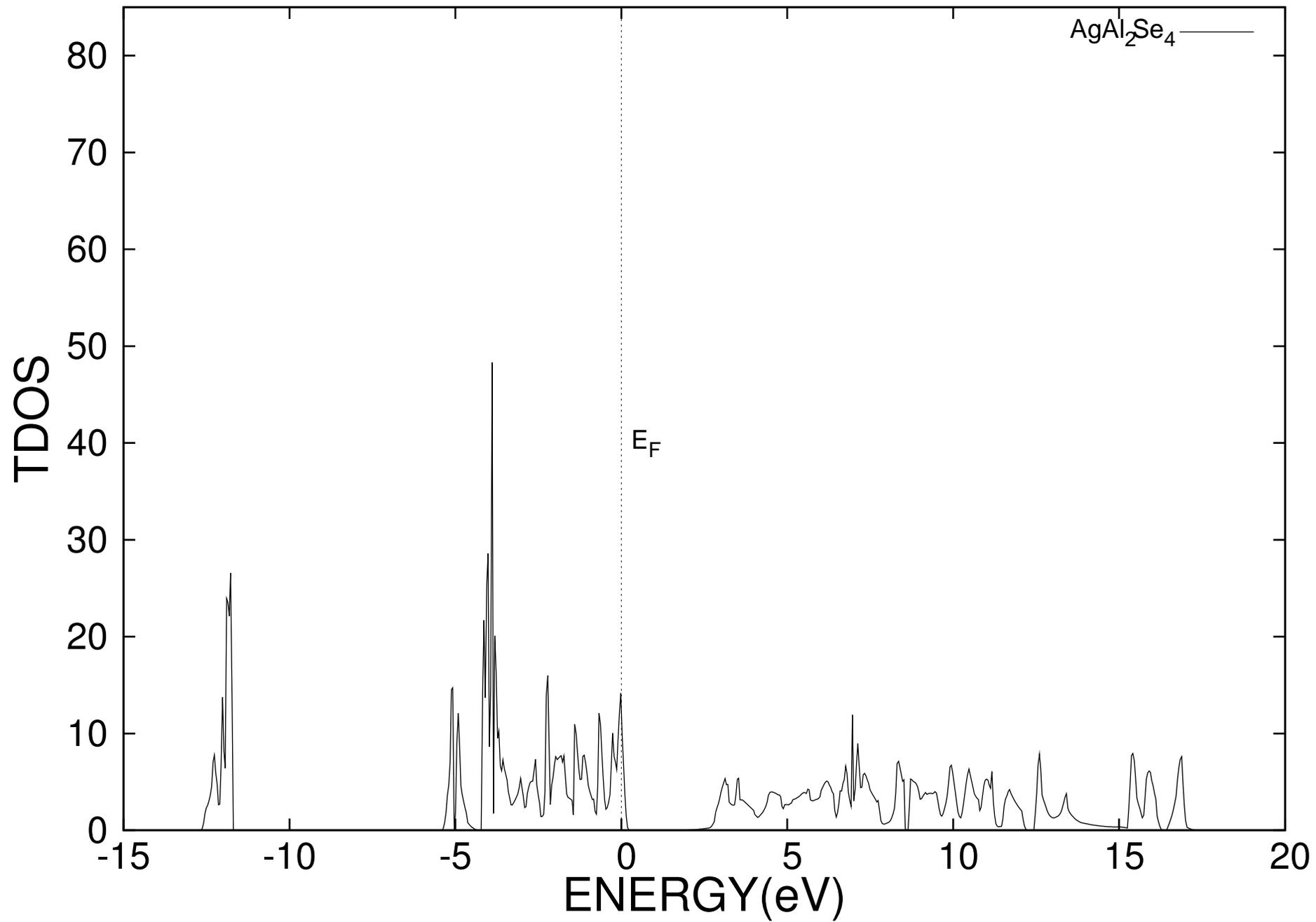